# Photonic Chiplet Interconnection via 3D-Nanoprinted Interposer

*Huiyu Huang[1, †], Zhitian Shi[1, †], Giuseppe Talli[2], Maxim Kuschnerov[2], Richard Penty[1], and Qixiang Cheng[1, *]*

[1] Centre for Photonic Systems, Electrical Engineering Division, Department of Engineering, University of Cambridge, Cambridge CB3 0FA, UK

[2] Huawei Technologies Duesseldolf GmbH, European Research Center, Riesstraße 25, 80992 Munich, Duesseldolf, Gemany

E-mail: qc223@cam.ac.uk

**Abstract**

Photonic integrated circuits (PICs) have been investigated using a variety of different waveguide materials and each excels in specific key metrics, such as efficient light emission, low propagation loss, high electro-optic efficiency, and potential for volume production. Despite sustained research, each platform shows inherent shortcomings that as a result stimulate studies in hybrid and heterogeneous integration technologies to create more powerful cross-platform devices. This is to combine the best properties of each platform; however, it requires dedicated development of special designs and additional fabrication processes for each different combination of material systems. In this work, we present a novel hybrid integration scheme that leverages a 3D-nanoprinted interposer to realize a photonic chiplet interconnection system. This method represents a generic solution that can readily couple between chips of any material system, with each fabricated on its own technology platform, and more importantly, with no change in the established process flow for the individual chips. Mode-size engineering is enhanced by the off-chip parabolic micro-reflectors. With a 3D-nanoprinted chip-coupling frame and fiber-guiding funnel, low-loss fully passive assembly and alignment can be achieved. A fast-printing process with sub-micron accuracy, achieving a mode-field-dimension (MFD) conversion ratio of 5:2 from fiber to chip is demonstrated with a low excess loss of <0.5 dB on top of the 1.7 dB inherent coupling loss. This is, to the best of our knowledge, the largest mode size conversion using non-waveguided components. Furthermore, we demonstrate such a photonic chiplet interconnection system between silicon and InP chips with a 2.5 dB die-to-die coupling loss, across a 140 nm wavelength range between 1480 nm to 1620 nm. This hybrid integration plan can bridge different waveguide materials, supporting a much more comprehensive cross-platform integration.

# 1. Introduction

PIC technology not only powers the modern internet, but is also widely acknowledged as a game-changing technology that promotes a host of innovations, including integrated quantum processing units [1,2], artificial intelligence accelerators [3,4], and light-based detection [5] and ranging imaging systems [6], etc. Arguably, the two most prominent integration platforms that have been widely commercialized are indium phosphide (InP) and silicon-on-insulator (SOI) [7,8]. The former offers light sources and amplifiers with good electro-optic performance, while the latter offers high-volume manufacturing capacity and high light confinement. One approach to combine the light emitting/amplifying capability of InP with the full scalability of SOI, advanced integration schemes, spanning flip-chip bonding [9], die/wafer bonding [10], micro-transfer printing [11], and direct epitaxial growth [12], have been heavily researched. Alternatively, to harness the potential of photonic integration platform to meet the ever-increasing circuit-level performance metrics, such as low waveguide propagation loss [13], high electro-optic coefficient [14], high Kerr coefficient [15], and second-order nonlinearity, etc., a range of different waveguide materials, such as silicon nitride [15], lithium niobate [10], gallium nitride [16], and aluminum nitride, have all been investigated. Foreseeably, to break the limitations on a single material platform, hybrid and heterogeneous integrations that create different cross-platform systems will continue to play a critical role. Indeed, recent demonstrations on III-V gain materials co-integrated on thin-film lithium niobate and silicon nitride [17] have opened new opportunities for high-performance chip-scale optical systems. From the recent demonstrations on cross-platform integration (**Supplementary I**), it is noticeable that the most important property of each material to enable the integration is probably its refractive index as it defines the achievable refractive index contrast to form a waveguide and hence the achievable waveguide mode diameter. Therefore, different integration technologies often come with different material interfacing methods, ranging from grating coupling, butt coupling, adiabatic coupling, to evanescent coupling. The primary considerations are generally low inter-waveguide coupling loss, high fabrication tolerance, and wide operation bandwidth [18]. The resulting required alignment tolerance is often regarded as key for scalable production capacity. Direct epitaxial growth of III/V materials onto the SOI platform may represent the ultimate path but currently has a low technology readiness level. Die/wafer bonding and micro-transfer printing are promising paths but require dedicated processing steps. Flip-chip bonding is a process generally performed after each die is fabricated on its own technology platform. A standard but additional process is required, i.e., back-end opening and metallization, to allow die placement and bonding.

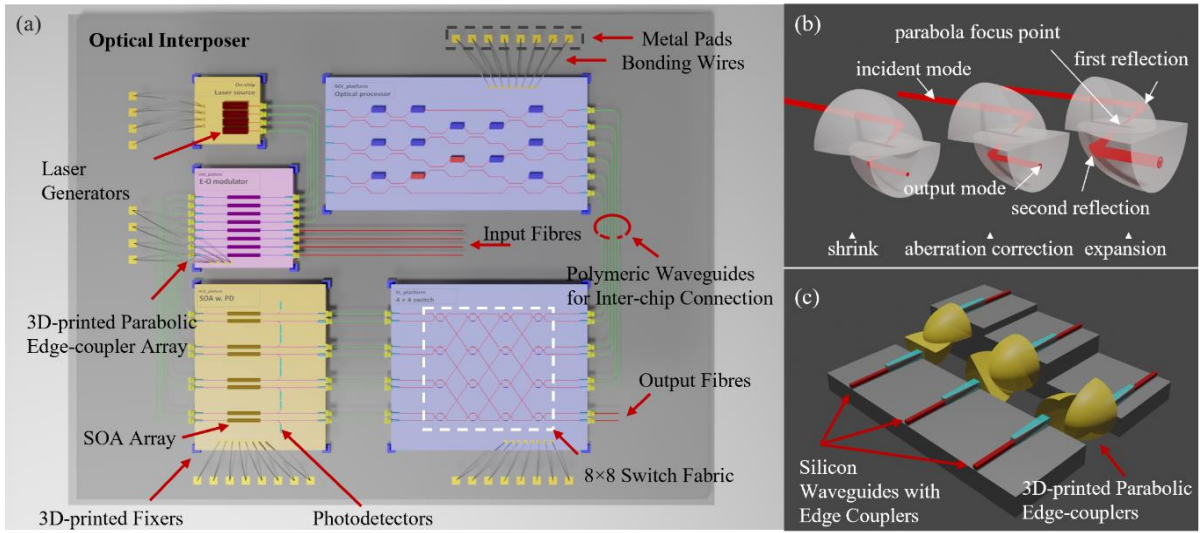

**Figure 1** (a) Conceptual schematic of chiplet interconnection realized by 3D printing. Notable components include on-chip laser generators and a semiconductor optical amplifier (SOA) array produced using a III/V platform, an electro-optic modulator chip fabricated with an LNO platform, and an optical processor and switch device from an SOI platform. The chiplets, originating from different optical platforms, are fixed by clamps onto an interposer. Mode size converters and aligning funnels are printed between different chips and fibers. (b) Schematic of the parabolic-shaped edge coupler. The incident beam propagates within the TPP resin structure and reflects at the resin/air interface. Based on the geometrical design of the edge coupler, the input Mode Field Diameter (MFD) can be maintained or expanded, and aberrations can be corrected by adjusting the relative positions of the two reflectors.; (c) Inter-chip couplers with arbitrary rotation angle. The couplers are directional-insensitive, such that they can be arranged to have a certain angle to help compensate height differences between adjacent chips.

Photonic wire bonding (PWB) stands out as an assembly technology for multi-platform photonic interconnection with great potential. It only requires coarsely pre-positioned photonic dies with the assistance of a 3D machine vision technique boasting sub-100 nm precision. Yet, additional post-processing steps, such as back-end opening, are imperative to expose the waveguide from the cladding layer. It is also noteworthy that the printed optical wires leave relatively substantial footprints on the chip, measuring at least a few tens of micrometers, as a prerequisite for achieving reasonable adiabatic coupling efficiency. The nature of this linear processing route inherently results in a time-consuming procedure.

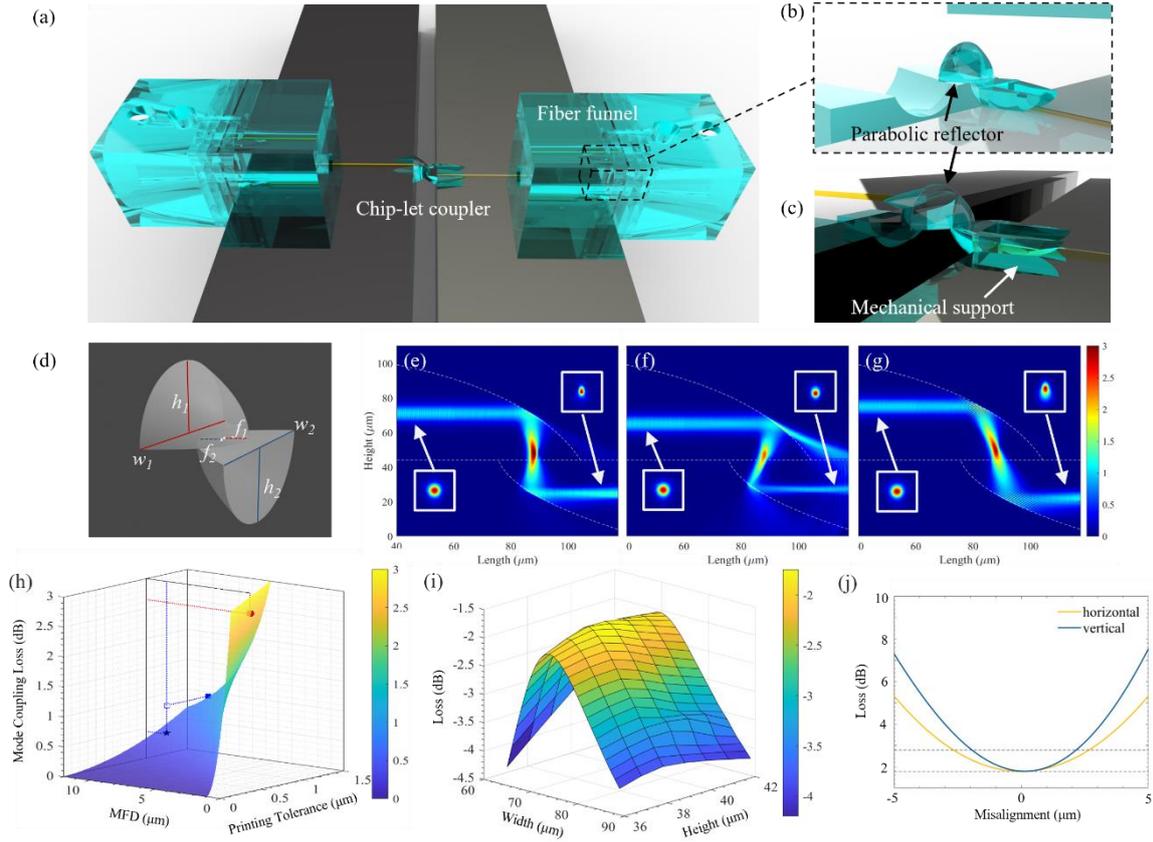

**Figure 2** (a) 3D rendering of a chip-to-chip testing sample. Three pairs of parabolic-shaped reflectors are placed at the input/output and inter-chip for fiber-to-chip and chip-to-chip coupling, respectively. The funnel-like structure on top of the chips is to accommodate optical fibers. Bayonet-shaped mechanical support is added to improve the stability. Close-up of (b) the fiber-to-chip coupling unit, and (c) the chip-to-chip coupling unit. (d) Schematic of a coupling unit consisting of a pair of parabolic-shaped reflectors, with key parameters highlighted. Light propagation paths simulated with Lumerical FDTD module at different laser injecting positions having (e) a balanced leakage and aberration, (f) too much leakage, and (g) too much aberration loss. (h) 3D plot of the mode coupling loss introduced by limited printing accuracy at different MFDs. The red solid dot and the blue star marks the targeted coupling loss for 4 μm and 10.4 μm MFD coupling, with 1 μm printing accuracy, respectively. (i) Mesh plot of the geometry optimization result from Lumerical FDTD simulations. (j) Parameter sweep results from the Lumerical FDTD simulation for the misalignment tolerance evaluation in horizontal (yellow) and vertical (blue) directions acquired at a wavelength of 1550 nm. The 1 dB-loss window is marked with the dashed lines.

In this manuscript, we present an alternative strategy for facilitating photonic chiplet interconnection via a novel 3D-nanoprinted interposer. The proposed optical interposer incorporates a substrate that accommodates 3D-nanoprinted elements responsible for establishing optical interconnections among dies of different material platforms and optical fibers. A conceptual hybrid chiplet integration is illustrated in **Figure 1a**, elucidating the spatial configuration of diverse 3D-printed components, including parabolic reflectors, mechanical supports, fiber funnels, and chip fixers. Its compact couplers that also handle arbitrary mode size conversion pave the way for high-density hybrid chiplet integration using 3D-nanoprinting (Figure 1b). The offset between the input and output optical axis can be utilized for height difference compensation by rotating the coupler (Figure 1c). By coordinating mechanical and optical structures, this co-design method ensures minimum requirement on chip alignment both spatially and angularly. Thus, it represents a generic solution that can readily couple between chips fabricated in material systems with zero change in their individual process flows.

We here demonstrate a micron-scale optical interposer that efficiently connects silicon and InP chips, with standard optical single-mode fibers (SMFs) as input/output connections. The arrangement of a simple two-chip integration is illustrated in **Figure 2a**, with close-ups of the fiber-to-chip and chip-to-chip coupling units plotted in Figure 2b and Figure 2c. This innovative photonic chiplet interconnection archives ultra-wide bandwidth, a substantial mode conversion ratio, and fully passive alignment capability.

## 2. Design and simulations

### 2.1. Basic concept of the parabolic coupling unit

The mode size converter employs a pair of parabolic-shaped reflectors. The incident light turns into a convergent beam after the first reflection at the resin/air interfaces, and travels through the focal point. The key of having a pair of parabolic-shaped reflectors is that, when they share the same focal point, the output beam turns back to a collimated beam that can be easily coupled to a waveguide. Furthermore, it enables efficient conversion of the mode field diameter (MFD), as detailed in the following subsection. The geometry of the parabola is tailored to precisely manipulate the light propagation. As illustrated in Figure 2d, the key parameters include the widths $w_1$, $w_2$, heights $h_1$, $h_2$, and the focal lengths $f_1$, $f_2$, of the two parabolic-shaped reflectors.

### 2.2. MFD conversion

When the pair of parabolic reflectors are symmetrical ($f_1 = f_2$), the input MFD will remain unchanged at the output. Changing the second reflector's dimensions alters the MFD of the output mode, achieving effective mode size conversion at low loss, as shown in Figure 1b.

Due to the free-space propagating nature of light inside the coupler and its 3D designing freedom, compact but ultra-wide bandwidth coupling can be achieved.

Considering a monochromatic, highly collimated incident laser beam, the output mode will have a new MFD $d_2$ of $d_1*f_2/f_1$, where d is the MFD of the input mode, $f_1$ and $f_2$ being the focal length of the first and second parabola, respectively. Thereafter, we can denote the MFD conversion ratio $\varphi$, as $\varphi = f_1/f_2$. Lumerical FDTD (finite difference time domain) software is subsequently used to visualize the light propagating inside the reflectors input from a guided mode and enables optimization of the design. An example is as follows: Figure 2e shows a side view of the optical propagation path at the wavelength of 1550 nm, and the parameters of the two parabolic-shaped reflectors are set differently to shrink the output mode. However, due to the polychromatic nature of the laser beams and the spherical aberration introduced by the reflections happened at the concave polymer/air interfaces, the geometry of reflectors needs to be tailored to optimize coupling efficiency (Figure 2f) and correct comatic aberration (Figure 2g). The focus point is not located at the interface of the two parts, and this is mainly because the input mode lose constrains from the core/cladding interface after it leaves the optical waveguide and becomes a divergent beam. The actual focal points of the reflectors are tweaked with the assistance of FDTD modeling. For the actual test, limited by the tool's alignment capability, the coupling loss will be slightly higher than the designed value, which is explained in Figure 2h. A more detailed description of the mode adaptation can be found in **Supplementary II**.

### 2.3. Geometry optimization

The geometry of the parabola is first calculated using linear optics to provide a course initial design with the target mode conversion ratio, taking the beam divergence into consideration. For the case of a Gaussian beam emitted at the end of the optic fiber that enters the reflecting units, the divergence half-angle, $\theta$, is calculated from:

$$\theta = \frac{\lambda}{\pi n w_0} \quad (1)$$

where $\lambda$ is the wavelength of the light source, $n$ is the refractive index of the resin that is used to shape the reflectors, and $w_0$ is the waist size. For the case of an integrated waveguide, similar estimations can be made.

Here, considering a laser beam that has a center wavelength of 1550 nm emitting from an SMF28 optical fiber that has an MFD of 10.4 µm, the divergence half-angle is calculated to be around 4°. Since the wavefront of the incident light is not planar, not all incident light is parallel to the main optical axis of the parabolic reflector. Therefore, comatic aberration occurs at the output. Also, the output mode will be expanded slightly due to the beam divergence. The incident angle at the output part then defines the output mode size, which can be described as:

$$2\left[\arctan\left(\frac{y_1 + t}{2x}\right) + \arctan\left(\frac{y_2 + t}{2x}\right)\right] \qquad (2)$$

where $y_1$ and $y_2$ denote the upper and lower boundary in the vertical direction, $t$ denotes the incident height and $x$ denotes the width of the parabola.

We simulate a pair of reflectors with Rhinocero for a proof-of-principle scenario with an SMF28 as input and an integrated waveguide of a 4 μm MFD as output. The model files are imported into Lumerical to conduct FDTD simulations. The input mode is set to have a central wavelength of 1550 nm, and the parameter sweep results of the reflector geometry are plotted in Figure 2i.

## 2.4. Aberration correction and leakage control

The reflectors are designed in such a way that they transmit the input with a low loss and with a high misalignment tolerance (Figure 2i and Figure 2j). However, geometrically, when the condition for Total internal reflection (TIR) is met, comatic aberration occurs, which is a major source to the signal loss as it results in some mode mismatch. To minimize the comatic aberration, a lower incident point is preferred for the input mode as shown in Figure 2f. However, this will in turn increase the loss due to light leakage as a result of the beam divergence during the free-space propagation within the reflectors that leads to a reduced incident angle at the resin/air interface. Therefore, a careful trade-off has to be made. The best incident spot and reflector geometry are identified using Lumerical FDTD with perfectly matched layer (PML) boundary conditions. The simulation result shows that 20 μm from the top of the parabola is the best position to balance light leakage and aberration. As a result of optimization via simulation, the mode size conversion loss can be minimized at around 0.7 dB, presumably bounded by the beam divergence. This could be further optimized by the use of irregularly shaped reflectors but this would compromise its ease of fabrication. The width and height of the reflector are optimized to be 82 μm and 38 μm, respectively, in this design example.

## 2.5. Alignment tolerance evaluation

Alignment tolerance is evaluated using Lumerical FDTD. The tolerance in both vertical and horizontal directions is greatly improved compared to the loss level without the parabolic reflectors. The 1 dB-loss window is increased from about 1 μm to about 6 μm in the horizontal direction and to about 4 μm in the vertical direction, as shown in Figure 2j.

To sustain a low coupling loss, it is required to keep both the input and output interfaces well aligned. We introduce a smart on-interposer coupling frame, composed of fixers and gripping clamps, as detailed in the following subsection, that is pre-printed to fix the position of the various dies. Subsequently, on-chip funnels are printed to ensure an accurate alignment of fibers. The coupling funnel's geometry is designed as 500 μm × 300 μm × 300 μm, truncated by a circular cone with a lower diameter of 130 μm and an upper diameter of 250 μm, taking resin shrinkage into consideration. The model of the frame intrudes into the chip edge by 20 μm, such that a tight fit between the resin frame and the chip edge is guaranteed after the printing process, and a sub-micrometer alignment accuracy can be achieved. A bayonet of 20 μm is presented at the end of the truncated cone to enhance the plug-in process's stability, as shown in Figure 2b.

## 2.6. 3D-nanoprinted chiplet fixing solution

The optical chiplet interposer concept is designed to leverage 3D-nanoprinted fixers to align chiplets and flexible clamps to provide additional grip force. **Figure 3** schematically describes the solution. The fixers are printed at the four corners where the chiplet will be placed, and the upper edges of the fixers are chamfered to allow the chip to be placed easily. A layer of resin under the fixer can be used to compensate for the height difference between different chiplets. The fixers enable a positioning accuracy of ±1 μm, and in order to push towards sub-micrometer accuracies, mechanical clamps are introduced, which are printed at opposite sides of a chiplet (demonstrated in **Section 4**). Besides, the fixers also help to ensure a smaller angular deviation between their respective axes of the chiplets, which eases the following on-chip printing aligning process. The crimps in the clamps that are pushed outwards by the chiplet provide a restoring force, which helps fix the chiplets in position and provides relative coordinates to help quickly locate and identify the markers for the following on-chip printing processes.

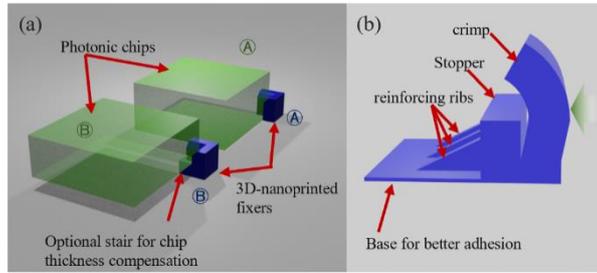

**Figure 3** Schematic of the 3D-nanoprinted chiplet fixing solution. (a) Fixers that can help compensate height differences, and (b) clamper for additional gripping force.

## 3. Fabrication and demonstration

### 3.1. Fabrication processes

The fabrication was carried out using a commercial direct laser writing system (Nanoscribe Professional GT2) with a printing setup consisting of a 25× objective and IP-n162 and IP-S resin provided by Nanoscribe. This choice of resin and objective lens is to ensure fast fabrication while maintaining good resolution. First, chip clamps and fixers are fabricated onto ITO (indium tin oxide) coated silica substrate to form the framework of the chiplet interposer. The chip clamps and fixers play a vital role in offering preliminary alignment. Then, dies of different materials are placed into the coupling frame formed by the clamps and fixers. After the chips are settled, the coupling funnels and parabolic mode size converters are fabricated onto the chips using one-step printing to ensure good alignment. The most critical step is the pre-printing alignment, which needs to be verified both vertically and horizontally. The relative coordinates between the chip edge and couplers offer a reference, whilst a finer adjustment is done horizontally with the assistance of the visual monitoring system in Nanoscribe GT2. As for the vertical aspect, we use the interface of the ITO-coated substrate and the base layer as a reference, taking advantage of their high refractive index contrast. The printing process utilizes a smaller laser power and slower scanning speed for the optical part to minimize surface scattering and a higher laser power and faster scanning speed for the mechanical part to reduce the printing time. The entire printing process takes approximately 25 minutes. After laser writing, the polymerized structure is developed in propylene glycol monomethyl ether acetate (PGMEA) for 15 minutes to remove unexposed resin, and in isopropanol (IPA) for 5 minutes to remove excess PGMEA. The device is then subjected to UV curing for improved mechanical stability.

## 3.2. MFD conversion

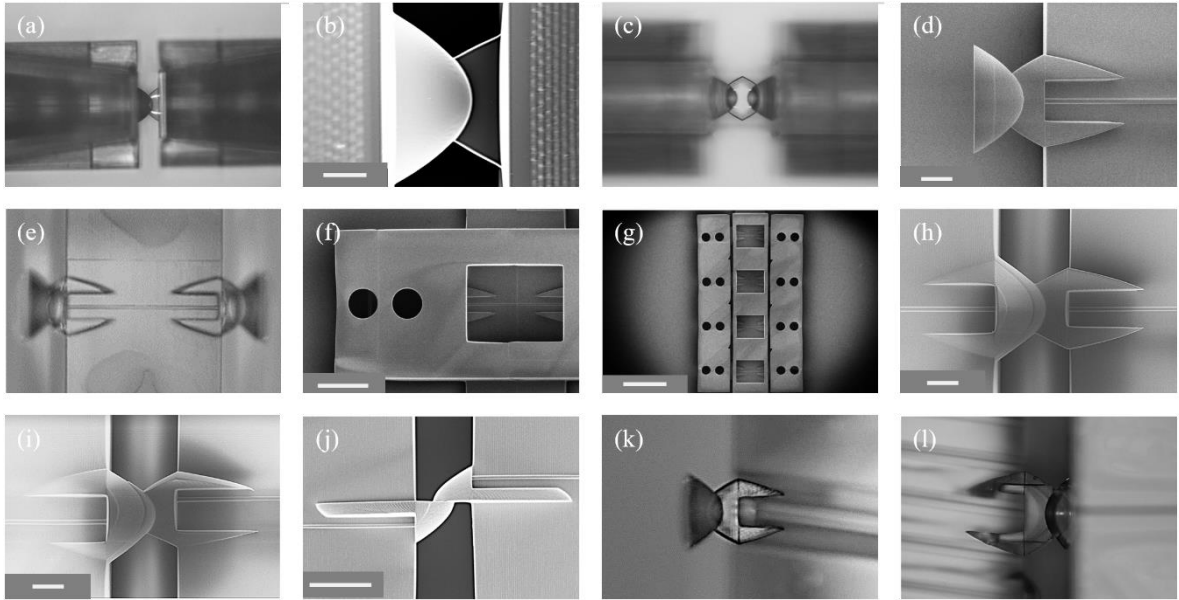

**Figure 4** (a) Microscope image of a fiber-to-fiber testing sample with two funnels at each side of the reflectors (scale bar 100 μm), and (b) SEM close-up of the reflectors, scale bar 20 μm. (c) Microscope image of a dual-pair reflectors for cascading validation, scale bar 100 μm. (d) SEM image of reflectors printed on a full-polymeric chip, scale bar 20 μm. (e) Microscope image (scale bar, 50 μm) of a fiber-to-chip test sample. (f) SEM image of the sample in, with the funnel structure added (scale bar, 60 μm). (g) SEM image of a four-channel fiber-to-chip test sample scale bar 300 μm. Close-up SEM image of chip-to-chip couplers with a rotation angle of (h) 0°, (i) 30° (h, i, scale bar 20 μm), and (j) 90° (scale bar, 40 μm). (k) Microscope image of an angled coupler (scale bar, 40 μm). (l) Microscope of an angled chip-to-chip coupler (scale bar, 40 μm).

In practice, the largest mode mismatch occurs for off-chip coupling where light couples from a standard single-mode fiber to a chip. As a result of the high index contrast of the silicon-on-insulator platform, the MFD in the edge coupler is about 5×7 μm$^2$. As a result, couplers with a mode diameter converting ratio as large as 2:1 are required as the MFD of SMF28 is about 10.4 μm, at a wavelength of 1550 nm. To evaluate the coupling and the MFD conversion loss, we use test couplers between optical fibers with different numerical apertures (**Figure 4a** and **4b**). The input mode has an MFD of 10.4 μm from SMF28, and the receiving optical fiber is selected to be either an SMF28 at a conversion ratio of 1:1, an HNA fiber that has an MFD of 6.8 μm and a conversion ratio of 3:2, or a UHNA fiber that has an MFD of 4.7 μm and a conversion ratio of 11:5. A coupler consisting of two pairs of reflectors is also developed for evaluating its cascading capability (Figure 4c) for higher conversion ratios. The excess coupling loss is, however, higher than expected due to the doubled non-constrained propagation of the divergent beam. For all testing, cleaved fibers are used. The fabricated prototypes are first attached to a 3-axis optical stage (Thorlabs MAX312D), and then characterized by a tunable laser (Keysight 81609A) sweeping from 1480 nm to 1620 nm.

## 3.3. Fiber-to-chip coupling

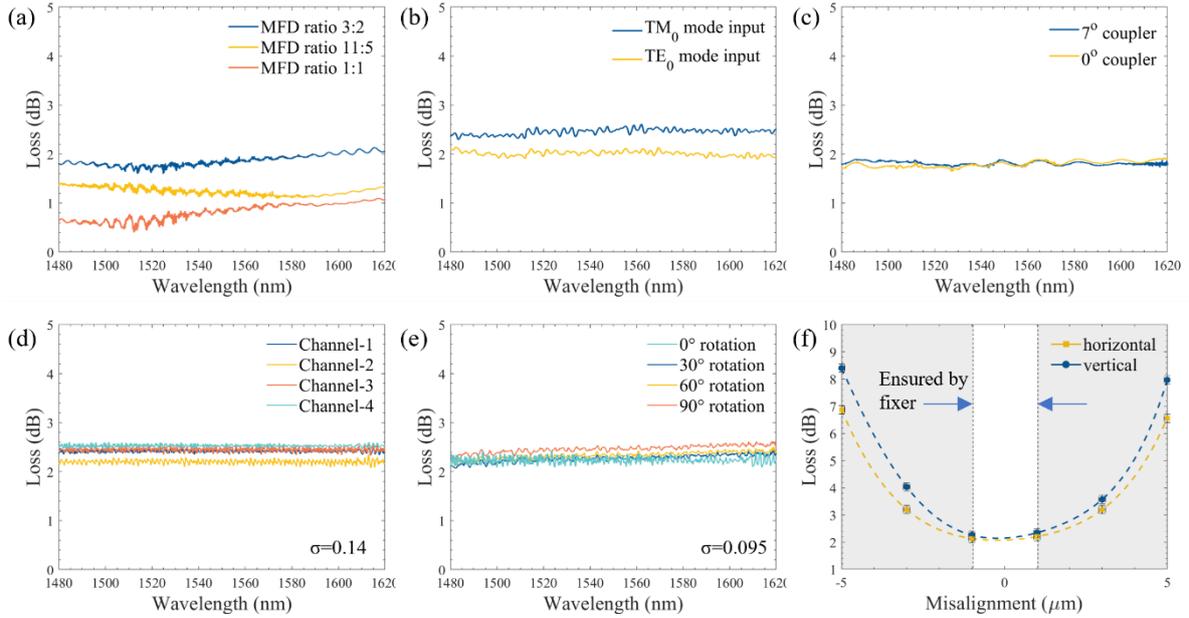

**Figure 5** (a) MFD conversion test with ratios of 1:1, 3:2 and 11:5. Measured coupling efficiency of (b) $TE_0$ and $TM_0$ modes, and (c) 7°-anti-reflection angled coupler. (d) Multi-channel fiber-to-chip coupling efficiency test. (e) Chip-to-chip coupling efficiency with the coupler rotated at an angle of 0°, 30°, 60°, and 90°. (f) Misalignment tolerance evaluation. A sub-micrometer accuracy can be achieved with the help of the funnel structure.

The fiber-to-chip coupling loss is evaluated by comparing the coupling efficiency with and without the presence of the coupler. The mechanical clamping unit is implemented for both cases to remove the effect of the alignment funnel on fibers. To better control variates, an in-house printed polymer chip using the IP-series resins with integrated waveguide is developed and used for a set of tests that help gain fresh insights into the tailoring of the coupling unit. A walkthrough of the sample chip preparation method is depicted in **supplementary Ⅳ**. Figure 4d, 4e, and 4f shows the printed test structures. The MFD of the polymer waveguide is designed to be around 4 μm which is close to that of the edge coupler in a foundry-produced Si chip. The parabolic-shaped couplers are printed on the polymer chip following the processes outlined in **Section 3.1**, to couple light into and out of the chip from the standard SMF28. This corresponds to a mode size mismatch of 84% difference in size. According to the Lumerical FDTD simulation, there is a theoretical loss of 1.7 dB/facet between the SMF28 and the chip, mainly coming from the mode mismatch. Experimentally, an input/output (I/O) loss of 2.4 dB/facet is measured over a wide range of wavelengths from 1480 nm to 1620 nm (see **Figure 5a**). Compared to the simulated coupling loss, the measurement result shows a total excess I/O-loss of 0.7 dB/facet that comes from the variations in the chip dimensions, surface roughness of the printed structures, reflections at the polymer/air and polymer/waveguide interfaces.

For photonic integrated devices, it is often desirable to achieve good coupling efficiency for both transverse electric (TE) and transverse magnetic (TM) modes. The fabricated coupler is thus also characterized using a polarimeter (Thorlabs PAX1000). The measurement results are plotted in Figure 5b. It is noticed that the $TM_0$ mode has a slightly higher loss than the $TE_0$ mode. This is mainly caused by the smaller effective index of the $TM_0$ mode, which will result in more leakage at the polymer-air interface. Besides, for the purpose of reducing reflection at the interface, a small tilted angle is usually introduced to the edge coupler. We tested both cases, with and without an anti-reflective angle of 7°. The measurement result is plotted in Figure 5c, and it shows that the coupler design has good compatibility with the angled edge couplers. Additionally, a common strategy to increase the optical I/O throughput is to have arrayed couplers. Thanks to the small footprint of the parabolic-shaped coupler, it is simple to have an array of couplers 3D-printed on-chip, as illustrated by Figure 4g. For now, the pitch is set at 300 μm, which is limited by the current funnel size to give the fiber robust support when plugged in. However, the pitch can be easily reduced as the adjacent funnels can be partly merged, without any compromise to mechanical stability. By optimizing the recipes, it is possible to reduce the pitch to 127 μm, a common standard used in photonic I/Os. Loss measurements for the four-channel prototype are shown in Figure 5d, with the loss consistent in all four channels.

### 3.4. Optical chiplet interconnecting

Similarly, polymeric chips that contain integrated waveguides are fabricated and used to emulate die-to-die interconnections between chiplets. The waveguides on the first and second chips are designed to have MFDs of 4 μm and 2 μm, respectively, in order to emulate typical MFDs of InP and Si waveguides. According to a Lumerical FDTD simulation, the coupling loss between the chips is 1.6 dB [19]. In the experiment, the chip-to-chip coupling loss is measured to be <2.5 dB, in which the excessive loss is believed to arise from the surface roughness but also the misalignment that is induced by the fabrication itself. The in-plane alignment of the chips is secured by the pre-printed clamps on the interposer. As photonic chips from different platforms often have different heights, it is necessary to compensate for the height difference whilst mode coupling. One of the advantages of the coupling unit is that, it can be rotated to help compensate for the height difference. In the experiment, the coupling unit is rotated 30°, 60° and 90° respectively, SEM images shown in Figure 4h, 4i and 4j. According to the tests, the coupling loss remains comparable with different rotation angles, with a loss of 2.3 dB (Figure 5e), which proves the ability of the device to solve the vertical alignment in chip-level coupling (Figure 5f).

## 4. Silicon-InP die-level interconnection

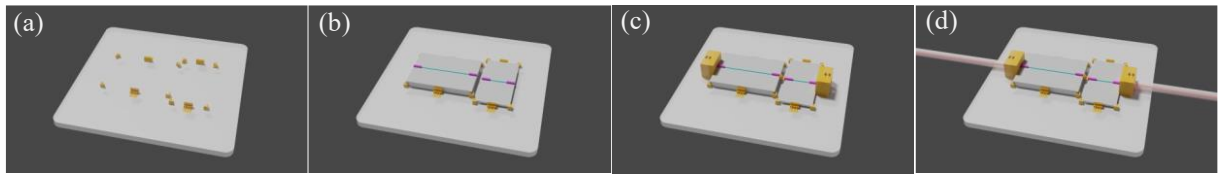

**Figure 6** Schematics of the process flow for the chiplet integration. The components are not in scale. (1) The fixers and clamps are printed on a silica substrate. Optional adhesives like crystalbond wax or UV glue can be applied. (b) The chiplets are placed with the help of fixers. (c) Reflectors and the funnels are on-chip-printed. (d) I/O optical fibers are inserted and guided by the funnel structures.

Finally, the developed designs are applied for silicon-InP die-level interconnection. Silicon and InP waveguide arrays are taped out using standard processes via multi-project wafer (MPW) runs, manufactured by Advanced Micro Foundry (AMF) and Heinrich-Hertz-Institute (HHI), respectively. The AMF and HHI chips are 2×3 mm$^2$ and 4×6 mm$^2$ in size, respectively, with a height difference of 525 µm. Both chips have an array of edge couplers as the I/O with a pitch of 200 µm. The AMF couplers are expected to have a mode cross-section of 7×9 µm$^2$, whilst the HHI ones are of 5×7 µm$^2$. The printed polymetric coupling structure itself provides a height compensation of 40 µm, and the rest is adjusted by the fixers with a polymer block underneath. The process flow for making the chip-let interposer is schematically shown in **Figure 6**. We introduce an anti-reflection angle to the edge couplers on both chips which is often needed on gain-integrated chips (Figure 4k). Two sets of fixers are printed on a silica substrate with relative positions to accommodate parabolic couplers, which form the framework of the optical interposer. Subsequently, the test chips are fixed onto the silica interposer, and the entire assembly is sent for a printing process wherein coupling units, including reflectors (Figure 4l) and funnels, are printed onto the chiplets. A microscope image of the silicon-InP co-integration sample with fibers plugged in is shown in **Figure 7a**, with close-ups of the key components. Figure 7b schematically explains how the chip height difference is compensated with the couplers and the fixers.

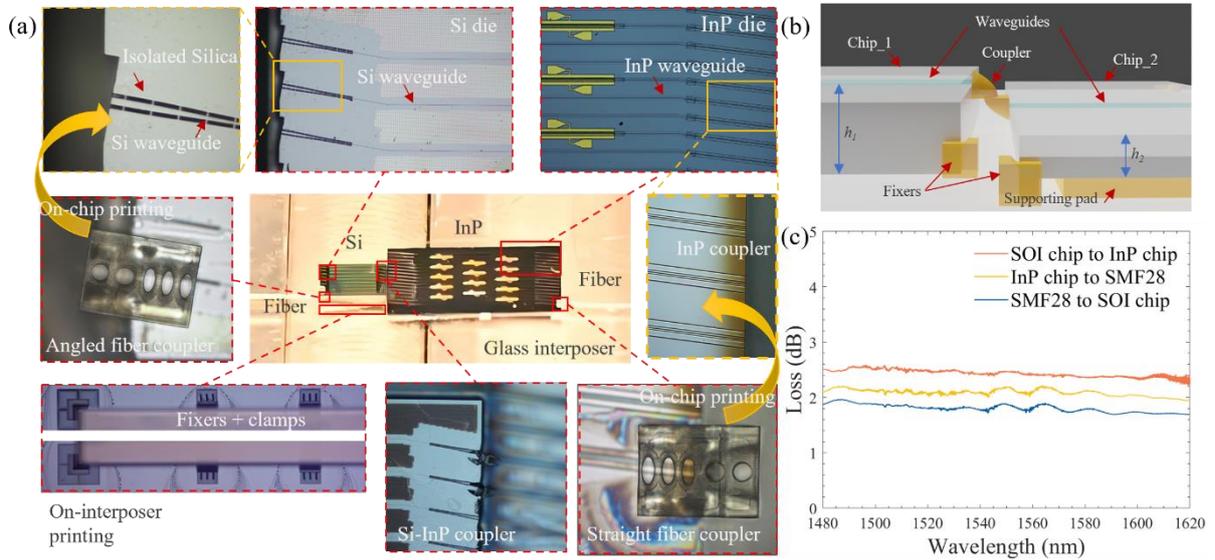

**Figure 7** Schematic of the coupling efficiency assessment for foundry-produced photonic chips. The SMF28 optical fiber, featuring a MFD of 10.4 µm, serves as the input path for the laser signal. The first coupler's funnel guides the fiber to the right position and adjusts the MFD to 7 µm to align with the silicon-based waveguide. Subsequently, the inter-chip coupler further shrinks the MFD to 5 µm, aligning with the III/V-based SOA array. The laser signal is then transmitted through the output fiber to a photodetector for measuring the output laser power, which is later used for the computation of coupler coupling loss.

For detailed characterization, the silicon and InP chips are assembled separately for the coupling loss breakdown. A tunable laser source (Keysight 81609A) sweeping from 1480 nm to 1620 nm is launched into the silicon, InP, and silicon-InP co-integration, respectively. After measuring the total loss from input fiber to output fiber via the two chips, the coupling efficiency is determined by subtracting the reference loss from both chips, and the detailed coupling loss is plotted in Figure 7c. The measured coupling loss between the Si and InP chips is around 2.5 dB. The coupling loss between the fiber and the Si, and the fiber and the InP are slightly below 2 dB and over 2 dB, respectively. The excess loss is again mainly due to limited fabrication accuracy and its resulting misalignment. By improving the printing recipes and voxel size, as well as the reflector shape, the coupling loss can be effectively limited to below 1 dB.

## 5. Outlooks and Conclusions

In this study, we introduce an adaptable yet powerful hybrid integration scheme that can interconnect photonic chiplets of different material systems, demonstrating both robust mechanical and optical performance. The pivotal contribution of our work lies in the demonstration of a compact photonic interposer that comprises both mode size converters and frames interconnecting waveguides of different material systems with zero change in established fabrication processes. This configuration is capable of ensuring passive alignment with minimal optical losses, despite current constraints in the fabrication accuracy.

This work so far focuses on photonic chip-let interconnections and the electrical connection aspect has not been demonstrated. However, in the envisioned system illustrated by **Figure 8a**, the electrical connections can be handled by the mature wire-bonding technique [20]. For a proper demonstration, this would require metallization on the glass interposer, which is a standard process but beyond our current capability.

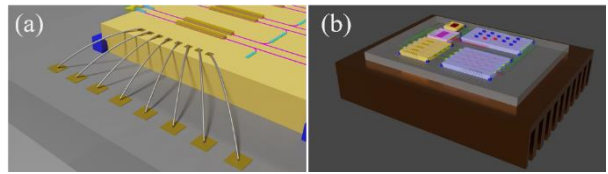

**Figure 8** Schematics of the future works. (a) Electrical connection with on-interposer metal contact pads. (b) Thermal management by using a heat sink as the substrate for the interposer.

Alternatively, we would also envision that existing flip-chip-bonding technology [21] can also be a good fit for our 3D-nanoprinted interposer solution to facilitate the fanout of high-density electrical connections. In this case, 3D-printing will be conducted on the substrate with dies flipped upside down, where visitable markers are needed. Another noteworthy consideration is thermal management. The incorporation of a heat sink or cooling pad is important, and the current silica substrates can be readily substituted with materials of high thermal conductivity, such as silicon. This allows direct mounting to a heat sink or cooling pad (Figure 8b).

The mode size converter, a fundamental component of our design, incorporates a pair of parabolic-shaped reflectors, which enables the unique capability of dynamically compressing or expanding the mode size with high coupling efficiency. To the best of our knowledge, we have achieved the largest mode conversion ratio of 5:2 using non-waveguided components. The versatility of the parabolic-shaped coupler is demonstrated with angled placement, twisted rotation, polarization-insensitivity, and arrayed implementation, all of which exhibit comparable loss levels. We have also successfully implemented a silicon-InP co-integration with plugged-in optical fibers using such an optical interposer. The excess loss mainly arises from the alignment variation due to the limited printing accuracy. This can be improved with a smaller voxel size of the two-photon polymerization (TPP) tool, which will in turn improve the structure surface roughness. In addition, the phase delay which is partially responsible for the coupling loss can be further reduced by fine-tuning the reflecting surface and making them 'free-form' [18]. To simplify the parabola design and facilitate a fast printing process, we stay with the perfect parabolic-shaped reflecting surface in this work. However, we foresee that coupling losses below 1 dB can be well achieved by making adequate improvements, making this hybrid integration approach hold great promise to stretch the limits of current PIC-level performance metrics without change of established process flows.

**Acknowledgements**

Huiyu Huang and Zhitian Shi contributed equally to this work. The authors would like to thank the Cambridge EEDB for providing the facilities and professional support. We thank Dr. Tian Gu from MIT for his helpul advice on the fabrication.

# Supporting Information

**Supplementary 1 Recent development of 3D-nanoprinted coupling methods for photonics applications**

3D-nanoprinting technology has shown its strength in photonic packaging via fabricating in-situ optical couplers for mode match, especially when dealing with special cases that involve non-standardized optical fibres or waveguides. In 2016, Dietrich et al. [1] demonstrated a convex-lens-based approach which shows a very low loss of 0.5 dB/facet using the active alignment method. Later [2], the same group demonstrated a series of lens-based couplers for connecting different platforms, all aligned dynamically. Gordillo et al. [3] printed an adiabatic-taper-based coupler with a funnel structure to enable passive alignment, which shows a good coupling efficiency of 2.5 dB/facet over a broad bandwidth of 100 nm. Blaicher et al. [4] presented their optical receiver using a convex lens to couple optical beam with a tapered coupler. This design is capable of mode size matching with a large conversion ratio. Yet, it is a one-way design that makes it hard to be used as an I/O port. Gehring et al. [5] introduced a very similar solution, which additionally has a reflecting surface at the backside of the convex lens, allowing a re-confinement of the light, making it a dual-way coupler. The coupler operates at the wavelength around 700 nm and the coupling efficiency drops quickly after 800 nm, far from the commonly used optical communication bands. In the same year, an optical-communication-compatible version was reported by the authors [6]. It is a very good design for single-chip applications, while the out-of-plane angle is not ideal for chip-to-chip coupling. Yu et al. [7] tried to couple the light with an on-chip printed freeform reflector. The reflector operates at wavelengths around 800 nm at a low coupling loss of about 1 dB/facet and is aligned actively. Gordillo et al. [8] published another manuscript on 3D-printed coupler that is equipped with a funnel-like structure to achieve passive alignment and stable fiber fixing. The impressive excess loss of 0.05 dB/facet proved that the mechanical fiber fixer is helpful to realize fully passive alignment, but the intrinsic grating coupler introduces a high coupling loss of 4.7 dB/facet with limited bandwidth. Chen et al. [9] demonstrated another one-way coupler, which is for light emitter. The coupling loss is reported to be 3 dB/facet and the 1 dB-loss-bandwidth is over 100 nm. In the same year, Luo et al. [10] published another approach which resembles Blaicher's work, still as a one-way coupler. Differently, Luo's model uses a reverse-pyramid-shaped structure as the light collector. In 2023, Yu et al. reported their improved version of the on-chip printed freeform lens array. It shows good off-plane coupling efficiency after the fiber array is actively aligned with the lenses. Xu et al. [7,11] reported a full-lens-based multi-platform integration in 2023. These demonstrates are summarized in **Table S1** with key figure of merits outlined. **Table S2** summarizes notable demonstrations of cross-platform integrations with different technologies.

Table S1. Recent development of 3D-nanoprinted coupling methods for photonics applications

| ref. | coupling type | emitter | receiver | loss (dB/facet) | alignment | mode size conversion ratio | broadband |
|---|---|---|---|---|---|---|---|
| [1] | lenses | SMF 10 µm | SMF 10 µm | 0.5 | dynamic | 1:1 | @ 1550 nm |
| [2] | lenses | SMF 10 µm | TriPleX 11 µm | 2.5 | dynamic | 0.9:1 | @ 1550 nm |
| [3] | direct | SMF 10 µm | polymer WG | 2.5 | passive | - | no change over 1520 nm -1620 nm |
| [4] | taper | SMF 10 µm | Si WG | 0.8 | dynamic | - | 1 dB loss over 100 nm (expected) |
| [5] | taper | S630-HP 4.2 µm | SiN WG | 3.1 | dynamic | - | @ 700 nm |
| [6] | taper | SMF 10 µm | SiN WG | 1.6 | dynamic | - | 1 dB over 1480 nm -1680 nm |
| [7] | EC | Polymer WG 2.3 µm | 780-HP 5 µm | 0.9 | dynamic | 0.5:1 | @ 856 nm |
| [8] | grating coupler | SMF 10 µm | Si WG | 0.05 | passive | - | @ 1544 nm |
| [9] | direct | Polymer WG 10 µm | multi-core fibers 10 µm | 3 | dynamic | 1:1 | 1 dB loss over 100 nm |
| [10] | taper | SMF 10 µm | SU-8 WG 3.5 µm | 1.04 | dynamic | 3:1 | 1 dB loss over 100 nm |
| [7] | EC | SMF 10 µm | SiN WG | 0.37 | dynamic | - | 1 dB loss over 500 nm |
| [11] | EC | SMF 10 µm | SiP-chip 10 µm | 1.44 | passive | 1:1 | 1 dB loss over 1475 nm-1625 nm |
| this work | EC | SMF 10 µm | polymer WG | 2.2 | passive | 5:2 | 1 dB loss over 1500 nm-1600 nm |

Table S2. Noticeable heterogeneous integration approaches for photonics packaging

| Ref | Material interfacing method | Integration technology | Coupling loss (dB/facet) | Material platforms | Bandwidth | Alignment |
|---|---|---|---|---|---|---|
| [12] | edge coupling | flip-chip bonding | 1.5 | InP and silicon wafer | @ 1550 nm | Active |
| [13] | evanescent coupling | wafer bonding | 12 | lithium niobates and III–V | @ 1580 nm | Lithograph |
| [14] | edge coupling | epitaxial growth | Not mentioned | 2H-MoTe2 on arbitrary interface | @ 1300 nm | Lithograph |
| [15] | adiabatic coupling | micro transfer printing | 12.3 | III–V and silicon | @ 1558 nm | Passive |
| [16] | edge coupling | photonic wire bonding | 4.2 | InP and silicon | @ 1493 nm | Active |
| This work | Edge coupling with mode size converting | Free-space coupler | 2.2 | multi-platform | 1480nm-1620nm | Passive |

# Supplementary 2 Requirements in mode size conversion

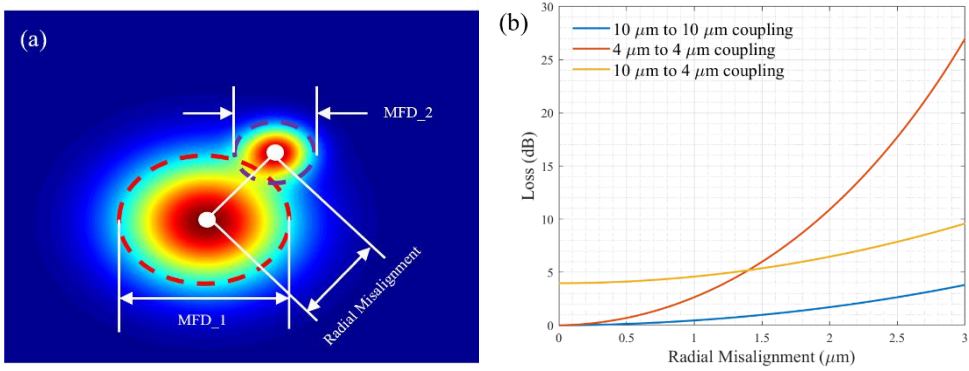

**Figure S1** Plot of the mode coupling loss introduced by limited printing accuracy at different MFDs. The coupling loss increases drastically with both misalignment increasing and MFD reducing. The red solid dot marks the designed coupling loss for 4 μm MFD coupling with 1 μm printing accuracy, and the blue star marks the coupling loss for 10.4 μm case. Experimentally, a lower printing accuracy introduces excess loss, making the actual loss higher than design target, shown as the solid/empty squares.

Truly passive alignment in optical coupling features a large tolerance. In instances where the Mode Field Diameters (MFDs) of paired modes are large, the aligning window is typically significantly expanded compared with smaller mode sizes (**Figure S1a**). Considering a near-Gaussian equilibrium mode distribution, the coupling efficiency, η, can be calculated as:

$$\eta = \frac{|\int E_1 * E_2 dA|^2}{\int |E_1|^2 dA \int |E_2|^2 dA} \quad (S1)$$

where $E_1$ and $E_2$ are the complex electric fields of the modes in the coupling plane, respectively.

The complex electric fields of the modes $E_1$ and $E_2$ are plotted in MATLAB. The MFD equals the diameter of the Gaussian beam where the intensity $I$, drops to the cut-off value of:

$$I = I_0 e^{-2} \quad (S2)$$

where $I_0$ is the peak intensity. Since the MFD $d$, is defined as two times of the spot size parameter $\omega$, which is proportional to the full width at half maximum (FWHM), then the MFD can be calculated with:

$$d = 2\omega = \frac{FWHM}{\sqrt{2ln2}} \quad (S3)$$

In Gaussian distribution, the FWHM equals to:

$$FWHM = 2\sqrt{2ln2}\sigma \quad (S4)$$

where $\sigma$ is the standard deviation. Hence, the MFD is denoted as:

$$d = 2\sigma \quad (S5)$$

Then, an offset is introduced to the centers of the two modes as the misalignment. Applying **Equation S1**, we can have the coupling efficiency calculated. In such a way, Figure 2i and Figure S1b are plotted. As illustrated in Figure S1(a), the relationship between MFD and alignment tolerance is elucidated. In Figure S1(b), for the case of a 10 μm-to-10 μm symmetrical mode coupling, the loss to misalignment curve is plotted as a solid blue line; a 4 μm-to-4 μm coupling case is plotted as a solid red line; and an asymmetrical case of 10 μm-to-4 μm coupling is plotted as a yellow solid line. From the plot, we know that the 10 μm-to-10 μm case has the largest 1 dB-loss misalignment tolerance. When the fiber/chip MFDs difference is large (ie. 10 μm vs. 4 μm), the 1 dB-loss misalignment tolerance is larger than 1 μm, and the loss mainly comes from the mode mismatch. When the 10 μm mode is converted to have a perfect match with the 4 μm mode, the misalignment becomes the dominating factor (case 4 μm-to-4 μm).

Functioning as the primary optical component in the system, the mode size converter incorporates a pair of parabolic-shaped reflectors. In this design, incident light from an input waveguide is directed into the initial segment of the coupling unit, undergoing two reflections at the polymer/air interfaces. Subsequently, the light is redirected through the second part and into a second waveguide. The parabolic shape is meticulously optimized to precisely manipulate the propagation of light within the reflector. When the two parabolic reflectors are symmetrical, the output MFD remains constant. Introducing variations to the dimensions of the second mode reflector facilitates the alteration of the MFD of the output mode, achieving effective mode size conversion. The inherent free space propagation of light within the coupler, coupled with its 3D design flexibility, allows for arbitrary and broad bandwidth coupling to be seamlessly attained.

**Supplementary 3 Requirements in mode size conversion**

Two-photon polymerization (2PP) technology, as a forefront in advanced manufacturing, utilizes unique nonlinear absorption of femtosecond laser pulses to precisely fabricate microstructures. Center to this technology is the meticulous recipe tuning, where adjustments in laser intensity, exposure time, and scanning speed determine the resolution, geometry precision, and mechanical properties of fabricated microstructures. More importantly, the printing quality of the parabolic lens, which has a large effect on light transmission efficiency.

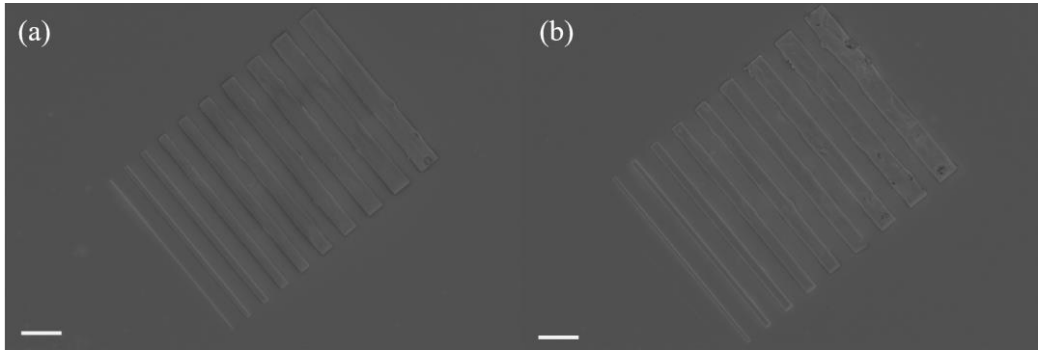

**Figure S2** SEM images of polymer waveguides with different printing recipe. (a) waveguides fabricated using optimized recipe (laser power 40 mW, scan speed 100000 µm/s), (b) waveguides use default recipe (laser power 100 mW, scan speed 100000 µm/s). All scale bars 20 µm.

In the fabrication process, the laser power and scan speed matter the most. Nanoscribe Photonics Professional GT2 is equipped with a femtosecond fiber laser with a center wavelength of 780 nm. The default laser power and scan speed of the laser are 50 mW and 100000 µm/s respectively, which provides good enough printing quality for general applications. However, as we are in a situation where any imperfection will lead to additional coupling loss, fine-tuning the printing recipes is critical. To achieve optimum performance, the laser power and scan speed must be carefully examined. In this work, a combination of 40 mW laser power and 100000 µm/s scan speed is adopted for a balance on printing time and accuracy. When printing on chips, the laser power is reduced by 5-10 mW subject to conditions. **Figure S2** are top-view SEM images taken from the dose test sample.

**Supplementary 4 Detailed alignment process of Si and InP chips**

The alignment procedure for the Si and InP chips involves an initial mechanical alignment process followed by further refinement using Nanoscribe. Mechanical fixers and clamps are first fabricated in accordance with the chip geometry. The gap between different fixers is set to be 5-10 μm lareger than the chip dimensions to allow sufficient room for chip insertion. Simultaneously, the supporting polymer pad is printed alongside the fixers and clamps in a one-step process using IP-S, ensuring high printing accuracy and alignment. A mechanical coupling frame is developed following Nanoscribe's standard procedure, and subesequently, the chips are inserted. For enhanced adhesion, wax and resins can be employed. While the mechanical coupling frame ensures assembly accuracy, achieving sub-micron alignment between optical interfaces proves still challenging.

Upon completion of the mechanical assembly, the entire structure, comprising chips, fixers, clamps, and polymer pad, is immersed in IP-n162 and sent to Nanoscribe for precise alignment. The built-in interface finder of Nanoscribe can accurately locates the chip-resin interface, facilitating the assessment of height differences between chips. Additionally, Nanoscribe's built-in camera aids in in-plane alignment by making the waveguide visible. Subtle rotations and adjustments of the parabolic-shaped structure enable the attainment of sub-micron-level alignment.

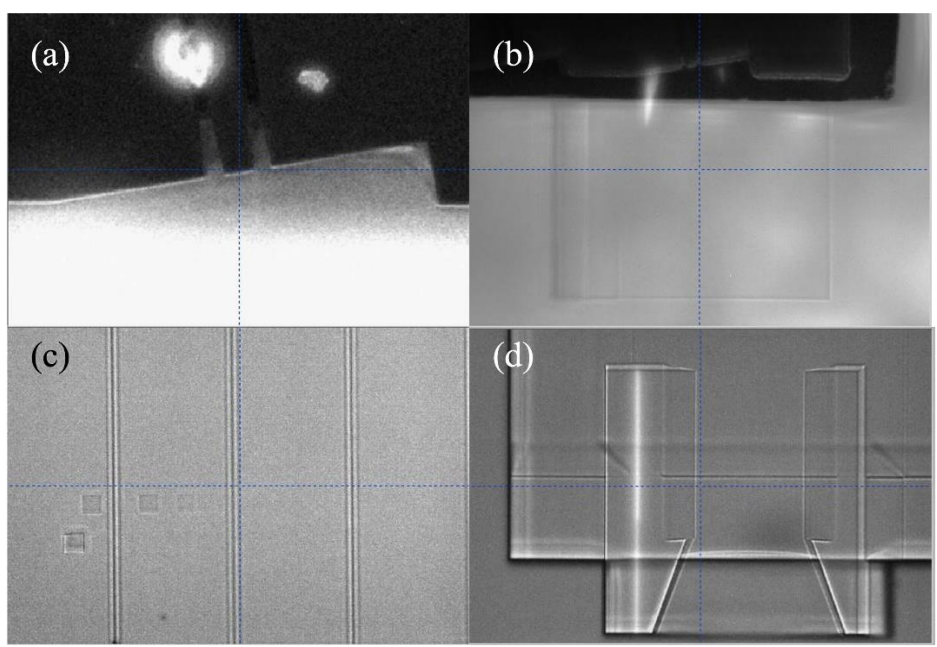

**Figure S3** Pictures from built-in camera of Nanoscribe. (a) center cursor helping in-plane alignment of printing, (b) checking object position while printing to ensure good alignment, (c) finding the interface position of the top layer, (d) overlap setting with built-in camera.

**Supplementary 5 Process flow for the fabrication of the full-polymeric test chips**

Detailed process flow for a full-polymeric test chip is depicted in the following figure. A block of resin is first printed on the carrier, which can either be a piece of glass or a silicon chip, and will later serve as the substrate of the whole testing structure. We use IP-S for the substrate, which is a photosensitive resin that has a relatively low refractive index provided by Nanoscribe. At the edge of the substrate, we introduce a set of align markers to facilitate the alignment process of the following steps. Then a polymer waveguide is printed using IP-n162, which has a higher refractive index than IP-S, to enable mode confinement. The parabolic-shaped reflectors are then printed at each end of the polymer waveguide, and precise alignment is ensured with the markers. Finally, the funnel-like structures are printed on the substrate to accommodate optical fibers during the coupling efficiency test. When creating the printing job for the funnels, some overlap with the substrate is needed to enhance the mechanical stability. The test results are presented in the main body of the paper.

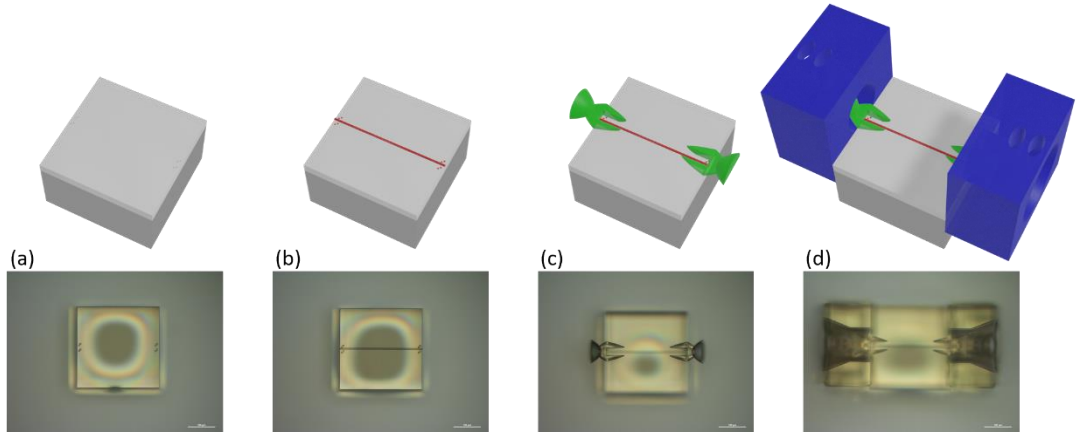

**Figure S4** Schematics (upper) and optical microscope images (lower) of the test sample preparation steps. (a) Print a base with IP-S resin which has a refractive index of 1.504 at the wavelength of 950 nm. The base layer helps to prohibit mode leaking during propagation. A pair of align markers are introduced at the edge of the base to enable coordination correction during the printing of the following structures. (b) The polymeric waveguide sits on the base. The photo-sensitive resin for the waveguide printing is IP-n162, which has a refractive index of 1.602 at the wavelength of 950 nm. (c) The parabolic-shaped reflectors are prepared at each end of the waveguide to couple the laser signal in/out of the test chip. The bayonet structure is split to avoid interaction with the waveguide. (d) The mechanical structure is printed over the reflectors to realize the passive alignment for the optical fibers.